\documentclass[12pt,aps,preprint]{revtex4} 
\usepackage{graphicx}
\usepackage{cancel}
\usepackage{amssymb}
\usepackage{color}
\def\beq{\begin{equation}}
\def\eeq{\end{equation}}                         
\def\bea{\begin{eqnarray}}
\def\eea{\end{eqnarray}} 
\def\to{\rightarrow}  

\begin{document}
\preprint{IPMU-08-0084,CAS-KITPC/ITP-073}
\title{Color Sextet Scalars at the CERN Large Hadron Collider}
\author{Chuan-Ren Chen~$^{\bf a}$}
\email{chuan-ren.chen@ipmu.jp}
\author{William Klemm~$^{\bf a,b}$}
\email{wklemm@berkeley.edu}
\author{Vikram Rentala~$^{\bf a,b}$}
\email{vikram.rentala@ipmu.jp}
\author{Kai Wang~$^{\bf a,c}$}
\email{kai.wang@ipmu.jp}
\affiliation{
$^{\bf a}$ Institute for the Physics and Mathematics of the Universe, University of Tokyo, Kashiwa, Chiba 277-8568, JAPAN\\
$^{\bf b}$ Department of Physics, University of California, Berkeley, CA 94720, USA\\
$^{\bf c}$ Kavli Institute for Theoretical Physics China, CAS, Beijing 100190, CHINA}

\begin{abstract}
Taking a phenomenological approach, we study a color sextet scalar at the LHC. 
We focus on the QCD production of a color sextet pair $\Phi_6\bar{\Phi}_{6}$ through $gg$ fusion and $q\bar{q}$ annihilation. 
Its unique coupling to $\bar{\psi^c}\psi$ allows the color sextet scalar to decay into same-sign diquark states, such as $\Phi_6\to tt/tt^*$. We  propose a new reconstruction in the multijet plus same sign dilepton with missing transverse energy samples ($bb+\ell^\pm\ell^\pm+\cancel{E}_T+Nj$, $N\geq 6$) to search for on-shell $tt\bar{t}\bar{t}$ final states from sextet scalar pair production. Thanks to the large QCD production, the search covers the sextet mass range up to  1 TeV for 100 fb$^{-1}$ integrated luminosity.  
\end{abstract}
\maketitle

\section{Introduction}
The Large Hadron Collider (LHC) at CERN will soon
provide a great opportunity for exploring physics 
at the TeV scale. As a proton-proton collider with a total center-of-mass energy of 14 TeV, the 
LHC is truly a Quantum Chromodynamics (QCD) machine. We therefore wish to study color
exotics, since any accessible new physics in the strong interaction sector will
 appear in the early stages of LHC operation. Many models of physics beyond the standard model (SM) naturally require the presence of color exotics, such as gluinos and squarks in supersymmetric 
extensions of standard model, $KK$-gluons and $KK$-quarks in extra dimensional models, or the top-prime in Little Higgs or twin Higgs models. All of these  are either quark or gluon partners 
which belong to the fundamental and adjoint representations of the 
QCD gauge group $SU(3)_C$ respectively. It is natural to consider colored particles in other representations;
in this paper, we focus on a scalar which is in the sextet(6) representation of $SU(3)_C$.  Color sextet particles have 
been widely discussed in nuclear physics as diquark condensate states; in the SSC era, sextet quarks 
were discussed in \cite{sextetquark}. 

Color sextet scalars are naturally present in partial unification \cite{pati}, grand unification \cite{pavel}  
and composite models; in some cases they may be present around the weak scale. 
For instance, in a supersymmetric Pati-Salam $SU(2)_R\times SU(2)_L\times SU(4)_C$ model,
light color sextet scalars can be realized around the weak scale, even though the scale of $SU(2)_R\times SU(4)_C$ symmetry breaking 
is around $10^{10}$ GeV due to existence of accidental symmetries with the masses of color sextet Higgs arising only through
high dimension operators \cite{Chacko:1998td,sextet}. In this case, 
the introduction of a color sextet Higgs will not lead to proton 
decay but only to neutron-anti-neutron ($n-\bar{n}$) oscillation and is fully compatible with present 
limits \cite{Chacko:1998td,sextet}. In a similar framework, light color sextet scalars also help in 
Post-Sphaleron baryogenesis \cite{babu}. In this paper, however, we will take a purely phenomenological approach 
toward the sextet scalar without assuming any model a priori. 

Among all the color exotics, the color sextet scalar is unique in its coupling to quarks.
In group theory language, the color sextet lies in ${3}\otimes {3} = {6}\oplus{\bar{3}}$ as a symmetric $2^{\rm nd}$ rank tensor under $SU(3)_C$. The Lorentz structure for this scalar coupling to quarks is given by $\psi^T C^{-1} \psi \phi$, where $\psi$ is a Dirac spinor and $\phi$ is the scalar. 
Under the SM gauge group $SU(3)_C\times SU(2)_L\times U(1)_Y$, the sextet scalar can be $\Delta_6$, a $SU(2)_L$ adjoint $(6,3,1/3)$; $\Phi_6$,
 a $SU(2)_L$ singlet $(6,1,4/3)$; $\phi_6$, a $SU(2)_L$ singlet $(6,1,-2/3)$; or $\delta_6$, a
$SU(2)_L$ singlet $(6,1,+1/3)$. The color sextet scalars are also charged under the global baryon symmetry $U(1)_B$ and 
the electromagnetic symmetry $U(1)_{\rm EM}$. To avoid breaking $U(1)_{\rm EM}$, these scalar fields should not develop a non-zero
vacuum expectation value. This condition removes any possibility of $n-\bar{n}$ oscillation in the minimal model involving color sextet scalars.
We may write down the flavor independent Lagrangian of such a minimal model by only considering SM gauge invariants and keeping $U(1)_{\rm EM}$
unbroken,   
\begin{eqnarray}
{\mathcal L} &=& \text{Tr}[(D_\mu \Delta_6)^\dagger (D^\mu \Delta_6)]-M^2_\Delta \text{Tr}[\Delta_6^\dagger \Delta_6]+ f_\Delta Q_{L} ^T C^{-1} \tau_2 \Delta^\dagger_{6} Q_{L}\nonumber\\
&+& (D_\mu \Phi_6)^\dagger (D^\mu \Phi_6)- M^2_\Phi \Phi_6^\dagger\Phi_6  + f_\Phi u^T_R C^{-1} u_R \Phi^\dagger_6\nonumber\\
&+& (D_\mu \phi_6)^\dagger (D^\mu \phi_6)- M^2_{\phi} {\phi_6}^\dagger\phi_6  + f_\phi d^T_R C^{-1} d_R {\phi}^\dagger_6\nonumber\\
&+& (D_\mu \delta_6)^\dagger (D^\mu \delta_6)- M^2_{\delta_6} {\delta_6}^\dagger\delta_6  + f_\delta d^T_R C^{-1} u_R \delta^\dagger_6\nonumber\\
&-& \lambda_\Delta(\text{Tr}[\Delta_6^\dagger \Delta_6])^2-\lambda_\Phi(\Phi_6^\dagger\Phi_6)^2- \lambda_\phi({\phi_6}^\dagger\phi_6)^2-\lambda_\delta ({\delta_6}^\dagger\delta_6)^2\nonumber\\
&-&\lambda^\prime_\Delta\text{Tr}[\Delta_6^\dagger \Delta_6 \Delta_6^\dagger \Delta_6]-\text{Tr}[\Delta_6^\dagger \Delta_6](\lambda_1{\Phi_6}^\dagger\Phi_6+\lambda_2{\phi_6}^\dagger\phi_6+\lambda_3 {\delta_6}^\dagger\delta_6)\nonumber\\
&-&\lambda_4 {\Phi_6}^\dagger\Phi_6 {\phi_6}^\dagger\phi_6 -\lambda_5 {\Phi_6}^\dagger\Phi_6 {\delta_6}^\dagger\delta_6 -\lambda_6 {\phi_6}^\dagger\phi_6 {\delta_6}^\dagger\delta_6~,
\end{eqnarray}
where the QCD covariant derivative is defined as $D_\mu = \partial_\mu -ig_s G^a_\mu T_r^a$, and the $T_r^a$ are the representation matrices for the sextet; $M^2_i$, $\lambda_i$ and $f_i$ are all positive-definite model parameters.

If we consider the $SU(2)_L$ adjoint sextet scalar $\Delta_6$ , there will be three physical sextet scalar states that couple 
to up-type quark pairs, down-type quark pairs, and up-down type quark pairs. When the sextet scalar decays into light quark states, 
the existing search strategies for massive octet scalars or 
vectors \cite{octet} may be employed. $\Delta_6$, $\Phi_6$ and $\delta_6$ may all contribute to the single top plus jet 
singal and $t\bar{t}+Nj$ signal from pair production. Here we consider the scenario in which a color sextet scalar decays into a
top quark pair so that one can use the leptons from the top quark decay to determine the features of the sextet. 
Signature that contains multi-top final states has been discussed in
the context of many new physics models as resonance decaying into top quarks or top composite\cite{sextet,top}.
To illustrate this and simplify our search, our study will focus on the color sextet $SU(2)$ singlet scalar $\Phi_6$ that only couples to righthanded 
up-type quarks.  

\section{Decay of the Color Sextet Scalar}

The decay of the $\Phi_6$ depends on its mass, $M_{\Phi_6}$, and its couplings to quarks, $f_{ij}~(i,j= u, c, t)$.  To illustrate our reconstruction algorithm in the discussion of discovery, 
we consider the case where $M_{\Phi_6}> 350$ GeV and the $\Phi_6$ decays into two onshell top quarks; other mass
ranges are discussed in the conclusion section. Above threshold, the general expression for the decay partial widths of the sextet scalar are 
\begin{eqnarray}
\Gamma_{ii} & = & {3\over 16\pi} |f_{ii}|^2 M_{\Phi_6} \lambda^{1/2}(1,r^2_i,r^2_i)(1-4r^2_i)\nonumber\\
\Gamma_{ij} & = & {3\over 8\pi} |f_{ij}|^2 M_{\Phi_6} \lambda^{1/2}(1,r^2_i,r^2_j)(1-r^2_i-r^2_j)
\end{eqnarray}
where $\lambda(x,y,z)=(x-y-z)^2-4 y z$ and $r_i = m_i/M_{\Phi_6}$.

By far, the most stringent bounds on these parameters come from $D^0-\bar{D}^0$ mixing, to which $\Phi_6$ would make a tree level contribution  proportional to $f_{11}f_{22}/M^2_{\Phi_6}$. The off-diagonal coupling $f_{ij}$ will contribute
to flavor violation processes, for instance $D\to \pi\pi$ which is proportional to $f_{12}f_{11}/M^2_{\Phi_6}$. The current bounds 
require that  
\beq
f_{11} f_{22} \lesssim 10^{-6}; f_{11} f_{12} \lesssim 10^{-2},
\eeq
for $M_{\Phi_6}$ of a few hundred GeV to TeV mass range \cite{sextet,Staric:2007dt,pdg}. One will also expect less stringent constraints from 
one loop process as $c\to u\gamma$. To escape from the bound, for acessible values of $M_{\Phi_6}$ we expect at least one of the couplings, $f_{11}$ or $f_{22}$, to be negligible. However from our purely phenomenological perspective, we take the decay branching 
fraction $\text{BR}(\Phi_6\to tt)$ to be a completely free parameter whose value may be determined at the LHC. 

\begin{figure}[!tb]
\includegraphics[scale=1,width=8cm]{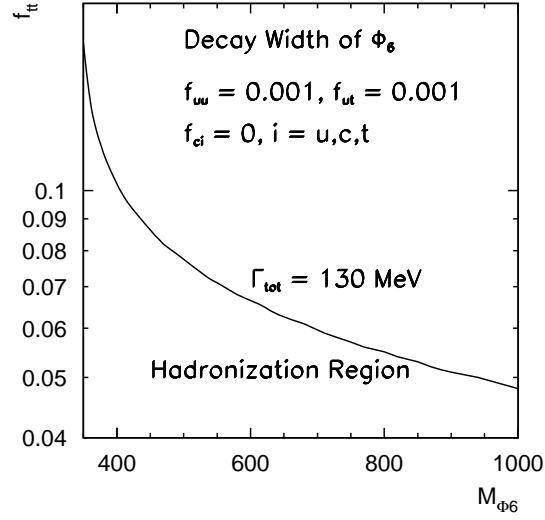}
\caption{Decay width contour for $\Phi_6$ in the mass and coupling plane.}
\label{decay}
\end{figure}

Because the sextet is a colored object, we need to consider the possibility of it hadronizing before decaying.  For example, it may form a 
tetraquark-like bound state with $\bar{3}\bar{3}$, such as $\Phi_6\bar{u}\bar{u}$, $\Phi_6\bar{u}\bar{d}$, or $\Phi_6\bar{d}\bar{d}$, with 
charges $0,1,2$ respectively. If the total width is less than $\Lambda_{\rm QCD}\sim m_\pi$, then the colored object will hadronize before it decays.  To determine the constraint imposed by the possibility of hadronization, in Fig. \ref{decay} we plot the contour for which decay width of $\Phi_6$ is equal to $\Lambda_{\rm QCD}$ as a function of the couplings and the mass.  Setting $f_{uu}=0.001$, $f_{ut}=0.001$, and eliminating any coupling to $c$, we see the possibility that a large portion of our parameter space will be protected from the risk of hadronization.

\section{Production of $\Phi_6$}

Because it carries color, $\Phi_6$ can be produced directly through the QCD strong interaction at the LHC.
The pair of $\bar{\Phi}_6\Phi_6$ is produced from gluon-gluon fusion or $q\bar{q}$ 
annihilation:  
\begin{eqnarray}
g(p_1) + g(p_2) \rightarrow \bar{\Phi}_6(k_1)+\Phi_6(k_2) \nonumber\\
q(p_1) + \bar{q}(p_2) \rightarrow \bar{\Phi}_6(k_1)+\Phi_6(k_2).
\label{eq:production}
\end{eqnarray}
The total production 
cross section depends only on the mass of $\Phi_6$, since the vertex is just the strong coupling, $g_s$, as shown in Eq. (\ref{eq:feynman_rule}). 
By comparison, the electroweak production of $\bar{\Phi}_6\Phi_6$ is small enough to be neglected in our search. 

From the scalar QCD gauge interaction
\beq
(D_\mu \Phi_6)^\dagger(D^\mu \Phi_6), ~~ \text{where}~~D_\mu = \partial_\mu -ig_s G^a_\mu T^a,
\eeq
one may obtain the Feynman rules
\begin{eqnarray}
G^a_\mu \Phi_6 \bar{\Phi}_6  &:& ig_s (p_1-p_2)_\mu T^a \nonumber\\
G^a_\mu G^b_\nu \Phi_6  \bar{\Phi}_6  &:& - ig^2_s g_{\mu\nu} (T^a T^b+T^bT^a) .
\label{eq:feynman_rule}
\end{eqnarray}
The momenta are assigned according to $V_\mu S(p_1)\bar{S}(p_2)$ and all momenta are out-going.
In group theory langauge, this is $6\otimes \bar{6} = 27 \oplus 8 \oplus1$.

The parton level cross sections for color a sextet pair production are given by
\beq
\sigma(q\bar{q}\rightarrow \bar{\Phi}_6\Phi_6)= \pi C(3)C(R){d_8\over d_3^2} {\alpha^2_s \over 3 s} \beta^3 = {10\pi\over 27 s}\alpha^2_s \beta^3
\eeq
and 
\begin{eqnarray}
\sigma(g g \rightarrow \bar{\Phi}_6\Phi_6) &=& d_R C_2(R) \pi{ \alpha^2_s \over 6s}{1\over d^2_8 }[ 3\beta(3-5\beta^2)-12C_2(R)\beta(\beta^2-2)\nonumber\\
&+& \text{ln} |{\beta+1\over \beta-1}|(6C_2(R)(\beta^4-1)-9(\beta^2-1)^2]\nonumber\\
&=& {5\pi\over 96 s} \alpha^2_s[\beta(89-55\beta^2)+\text{ln} |{\beta+1\over \beta-1}|(11\beta^4 + 18 \beta^2 -29)]~,
\end{eqnarray}
where $\sqrt{s}$ is the total energy, $\beta=\sqrt{1-4M^2_{\Phi_6}/s}$ and $R$ is $6$ with the normalization 
factor $C$ and Casimir $C_2$ satisfying 
\beq
\text{Tr}[T^a_R T^b_R]=C(R)\delta^{ab}~~\text{and}~~T^a_R T^a_R = C_2(R){\mathbf 1}.
\eeq
We list the values for different representations under $SU(3)$ as in Table \ref{tab:grp1}.
\begin{table}[!tb]
\begin{tabular}{| c | c c c |}
\hline
 $d_R$     &   $3$   &  $6$  &    $8$   \\
 $C(R)$    &   $1/2$ & $5/2$ & $3$\\
 $C_2(R)$  &   $4/3$ & $10/3$ & $3$ \\
\hline
\end{tabular}
\caption{Normalization factor $C(R)$ and quadratic Casimir $C_2(R)$ for $d_R=3,6,8$ under $SU(3)$.}
\label{tab:grp1}
\end{table}

\begin{figure}[!tb]
\includegraphics[scale=1,width=8cm]{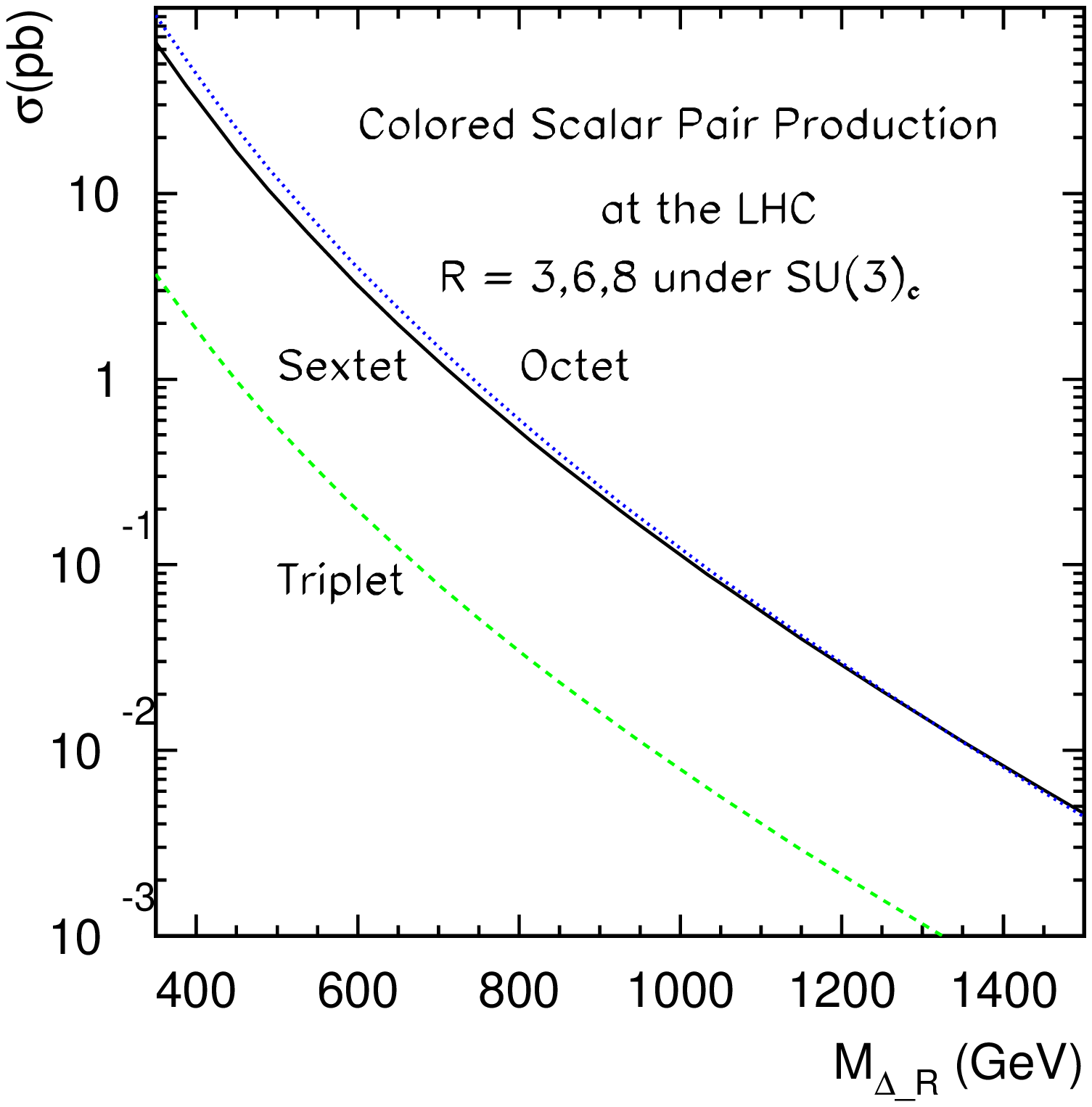}
\includegraphics[scale=1,width=8cm]{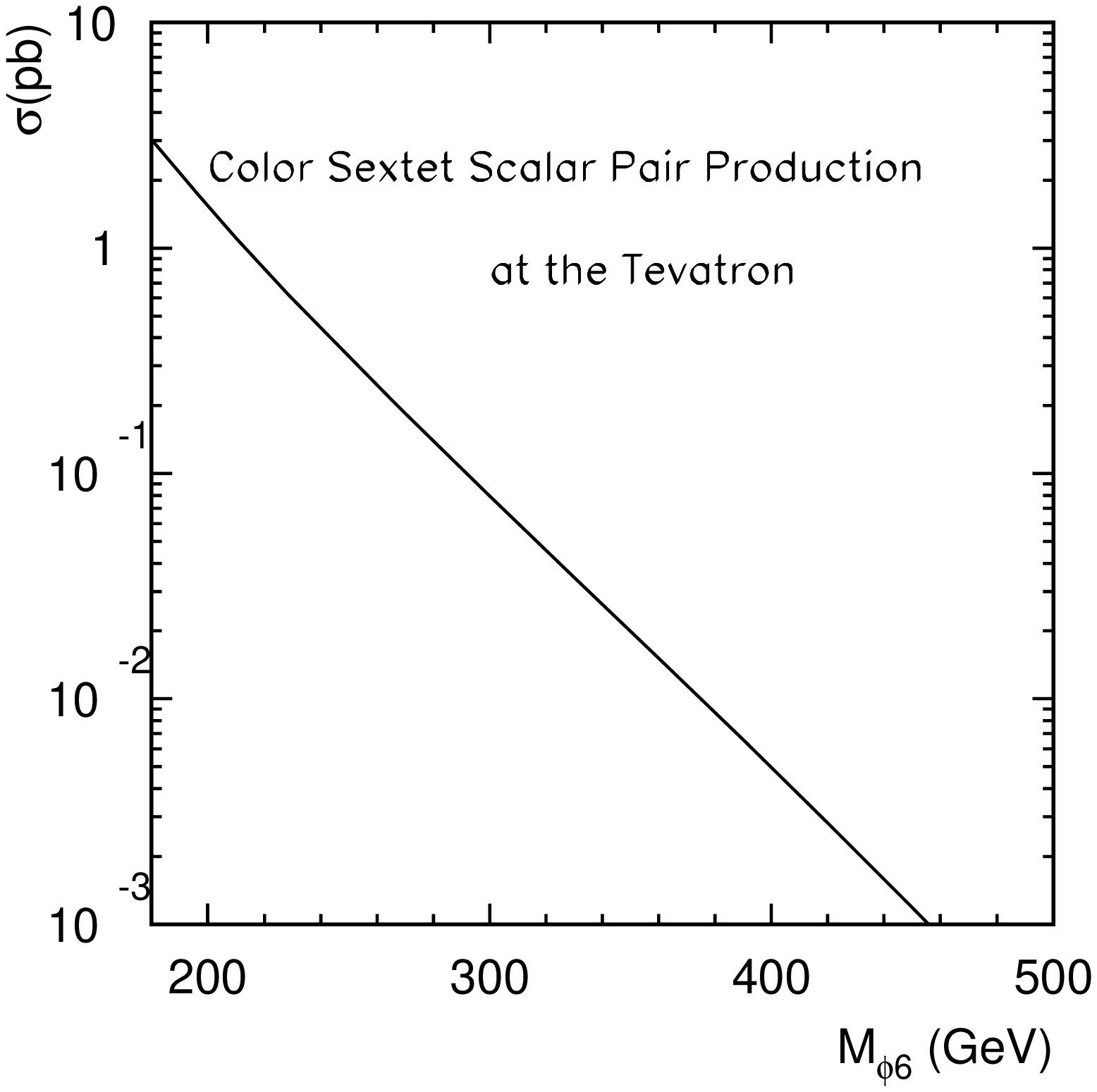}
\caption{Production of $\bar{\Phi}_6 \Phi_6$ at the LHC and Tevatron with $\mu_F= M_{\Phi_6}$, fixed scale $\alpha_S(\mu_R)$ with $\mu_R = m_Z$. 
The PDF set CTEQ6L has been used in all calculation. }
\label{total}
\end{figure}

The QCD production cross sections for the color sextet scalar pair $\bar{\Phi}_6\Phi_6$ at both LHC and Tevatron 
are plotted in Fig. \ref{total} with factorization scale $\mu_F= M_{\Phi_6}$, renormalization 
scale $\mu_R = m_Z$ and the CTEQ6L~\cite{Pumplin:2002vw} parton distribution function (PDF).
The matrix elements in our calculations here and elsewhere are generated 
by SUSY-Madgraph \cite{Cho:2006sx} with modified color factors. 
For comparison, we also show the pair production cross sections for $SU(3)_C$ triplet and octet 
scalars at the LHC. As we can see, the total production cross section of the sextet scalar is 
similar to that of the octet scalar, but is about one order maginitude larger than that of the triplet 
scalar, which can be understood from values of $C$ and $C_2$ for different reprensentations in Table~\ref{tab:grp1}.

As discussed in the introduction, the color sextet scalar $\Phi_6$ only couples to the righthanded up-type quark quark pair. 
Thus we may also have single production of a $\Phi_6$ through
\beq
uu (cc)\rightarrow \Phi_6.
\eeq
However, the production cross section is proportional to the coupling $|f_{uu}|^2$ and $|f_{cc}|^2$, and may therefore be suppressed due to the $D^0-\bar{D}^0$ mixing constraint. Some studies of the single $\Phi_6$ 
production at the Tevatron and the LHC have been done in Ref. \cite{sextet}. 

\section{Searching for the color sextet scalar through $tt\bar{t}\bar{t}$}

As discussed in the previous sections, the most distinct feature of the color sextet scalar is its decay mode $\Phi_6\to tt$, which leads to a same-sign dilepton signature in the final state if both top quarks decay 
semileptonically, i.e. $t\to W^+ b \to \ell^+ \nu b$. 
To avoid ambiguities in lepton assignments during reconstruction, 
we require the anti-top quark pair from the $\bar{\Phi}_6$ to decay hadronically. Hence, the final state of $\bar{\Phi}_6\Phi_6$ is
\beq
pp\to \bar{\Phi}_6\Phi_6\to tt\bar{t}\bar{t}\to 4b+\ell^\pm\ell^\pm+\cancel{E}_T+Nj, 
\eeq
where $\ell = e ~\rm{and}~ \mu$ and $N\geq 4$ to allow initial and final state QCD radiation. In our study, however,
the QCD radiation is not included.
To get this final state, the decay branching ratio will be
\beq
\text{BR} = \text{BR}^2(\Phi_6\to tt)\times \left({2\over 9}\right)^2\times \left({2\over 3}\right)^2\times 2,
\eeq
where the situation that top quark decays hadronically and anti-top quark decays semileptonically is 
also included. 
Figure \ref{total} also clearly shows that a color sextet with $M_{\Phi_6}\geq 350$ GeV will not be bounded by Tevatron data as the same-sign dilepton plus 
multi-jet final state from $tt\bar{tt}$ will be less than one event for $2~\text{fb}^{-1}$ luminosity.  
  
\begin{figure}[!tb]
\includegraphics[scale=1,width=8cm]{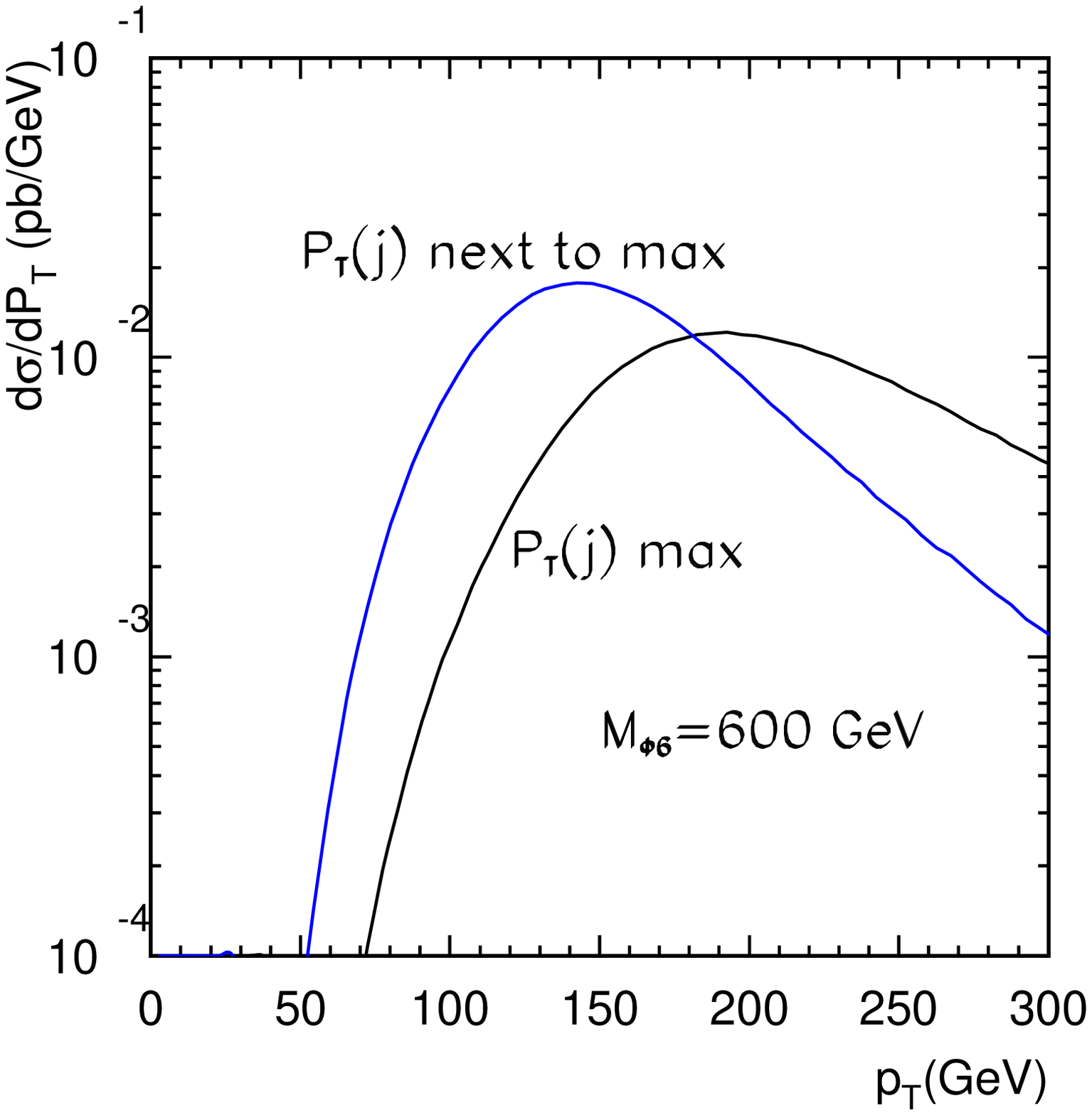}
\caption{$\text{max}\{p^J_T\}$ and $\text{next-to-max}\{p^J_T\}$ }
\label{jpt}
\end{figure}

To illustrate the kinematic features of the color sextet scalar pair, we consider the decay process
$\Phi_6 \bar{\Phi}_6\to tt\bar{t}\bar{t}\to bb\bar{b}\bar{b} \ell^+ \ell^{'+} + 4 jets$ and take $M_{\Phi_6} = 600$ GeV. 
The leading and second-leading jet $p_T$ distributions are shown in Fig. \ref{jpt}.  The typical hardness of these jets is the basis for one of our selection cuts introduced later in this section.  In order simulate the detector 
effects on the energy-momentum measurements, we smear the electromagnetic energy
and the muon momentum by a Gaussian distribution whose width is parameterized as \cite{CMS}
\begin{eqnarray}
{ \Delta E\over E} &=& {a_{cal} \over \sqrt{E/{\rm GeV}} } \oplus b_{cal}, \quad
a_{cal}=5\%, b_{cal}=0.55\% ,
\label{ecal}\\
{\Delta p_T\over p_T} &=& {a_{track}p_T \over {\rm TeV}} \oplus {b_{track}\over \sqrt{\sin{\theta}} }, \quad
 a_{track}= 15\%,b_{track} =0.5\%.
\end{eqnarray}
The jet energies are also smeared using the same Gaussian formula as in Eq.~(\ref{ecal}),
but with  \cite{CMS}
\begin{equation}
a_{cal}=100\%,\quad  b_{cal}=5\%.
\end{equation}

\begin{figure}[!tb]
\includegraphics[scale=1,width=8cm]{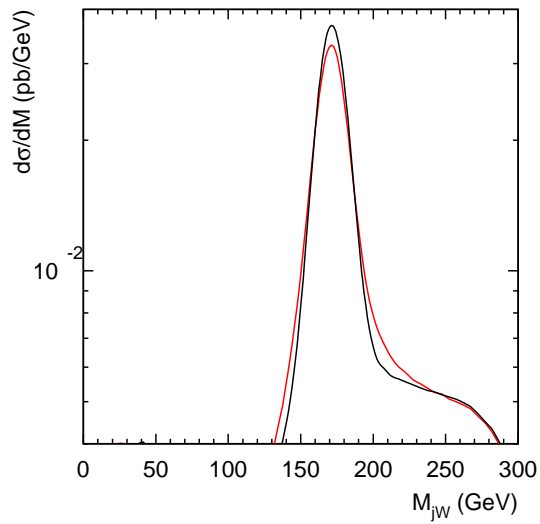}
\caption{Reconstructed Hadronic Top Pair.  The black(red) line represents the first(second) reconstructed hadronically decaying anti-top quark.}
\label{mbjj}
\end{figure}

We first reconstruct the two on-shell hadronically decaying $W$'s. Our procedure is to consider all dijet invariant masses except for those containing
one of the two tagged b-jets, since we require b-tagging in the event selection 
discussed later, and 
choose the two closest $M_{jj}$ combinations, which we then require to lie within the mass window
\beq
|M_{jj} - m_W| < 15 \text{GeV}.
\eeq
From this, we get the two reconstructed $W$ momenta. We then consider all combinations of reconstructed $p_W$ with 
all jets and again 
choose the two closest invariant masses
$M_{jW}$. In this way, we reconstruct the two hadronically decaying anti-top quarks.  
The distributions of these reconstructed invariant masses are shown in Fig. \ref{mbjj}.

Once we have the reconstructed two anti-top quarks, the reconstruction of the 
sextet $(\bar{\Phi})$
can be done using the 6-jet invariant mass $M_{6j}$ for the two hadronic anti-top quarks.  
Although the production of
neutrinos prevents us from fully reconstructing the sextet which produces the leptonic decays, 
we may reconstruct the transverse mass $M_T$ for the remaining two jets plus same-sign 
dilepton and $\cancel{E}_T$ as 
\beq
M_T=\sqrt{(\sum_{j} E_T+\sum_{\ell}E_T+\cancel{E}_T)^2-(\sum\vec{p}(j)+\sum\vec{p}(\ell)+\cancel{\vec{p}})^2_T} .
\eeq
\begin{figure}[!tb]
\includegraphics[scale=1,width=8cm]{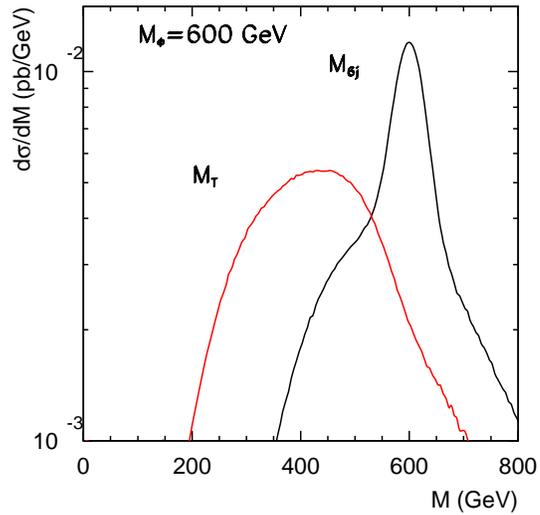}
\caption{Reconstructed Sextet from $m_{6j}$ and $M_T$.}
\label{m6j}
\end{figure}
As seen in Fig. \ref{m6j}, our reconstruction shows a clear resonance in both the $M_{6j}$ and $M_T$ distributions.

\begin{figure}[!tb]
\includegraphics[scale=1,width=8cm]{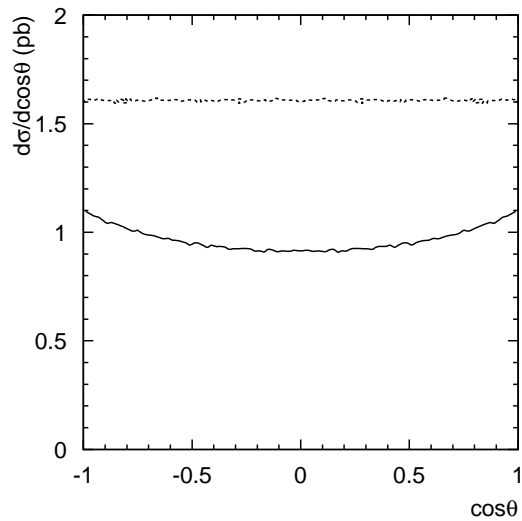}
\caption{Distribution of $\cos{\theta}$ between reconstructed top momentum and reconstructed sextet momentum. Dashed(Solid) line shows
the distribution without(with) smearing effects and kinematic cuts. }
\label{cos}
\end{figure}
Finally, since the two anti-top quarks may be fully reconstructed, we can 
boost back to the rest frame of the $\bar{\Phi}_6$ and 
study its spin. As shown in Fig. \ref{cos}, the angular distribution of the anti-quark clearly shows 
that the $\bar{\Phi}_6$ is a scalar. Since there are two missing neutrinos from decay of two top quarks, 
it is challenging to fully reconstruct top quark's momentum, and study the spin information of top quarks, 
which can be used to check this model since $\Phi_6$ only decays into a right-handed top quark pair. 
We leave this for future work. 

We next consider the backgrounds for our signal.  We require at least 2 tagged b-jets
 plus a same-sign dilepton and multijet. The irreducible SM background for this final state 
consists of $t\bar{t}W^\pm+Nj$, $bb+W^\pm W^\pm+Nj$ and $t\bar{t}t\bar{t}$. We 
estimate the QCD $bb+W^\pm W^\pm+Nj$ background by
computing $jjW^\pm W^\pm$ production. 
This is only 14 fb, and one expects the $bb+W^\pm W^\pm+Nj$ is about
three orders lower and therefore $< 0.1$ fb. The SM 4-top $t\bar{t}t\bar{t}$ is less than 0.1 fb to start with. The leading background
thus comes from $t\bar{t}W^\pm$ with one hadronic top decay and one semileptonic top decay with the same sign as $W^\pm$
leptonic decay. 

We propose the following selection cuts:
\begin{itemize}
\item $\text{min}\{p_T(j)\}> 15$ GeV, $\text{max}\{p_T(j)\}> 100$ GeV, $\text{next-to-max}\{p_T(j)\}> 75$ GeV, $|\eta(j)<3.0|$
\item same sign dilepton with $p_T(\ell)> 15$ GeV, $|\eta(\ell)<2.8|$
\item $\Delta R_{jj},~\Delta R_{jl},~\Delta R_{ll}> 0.4$
\item at least two b-tagged jets
\item $\cancel{E}_T> 25$ GeV
\end{itemize}

\begin{figure}[!tb]
\includegraphics[scale=1,width=8cm]{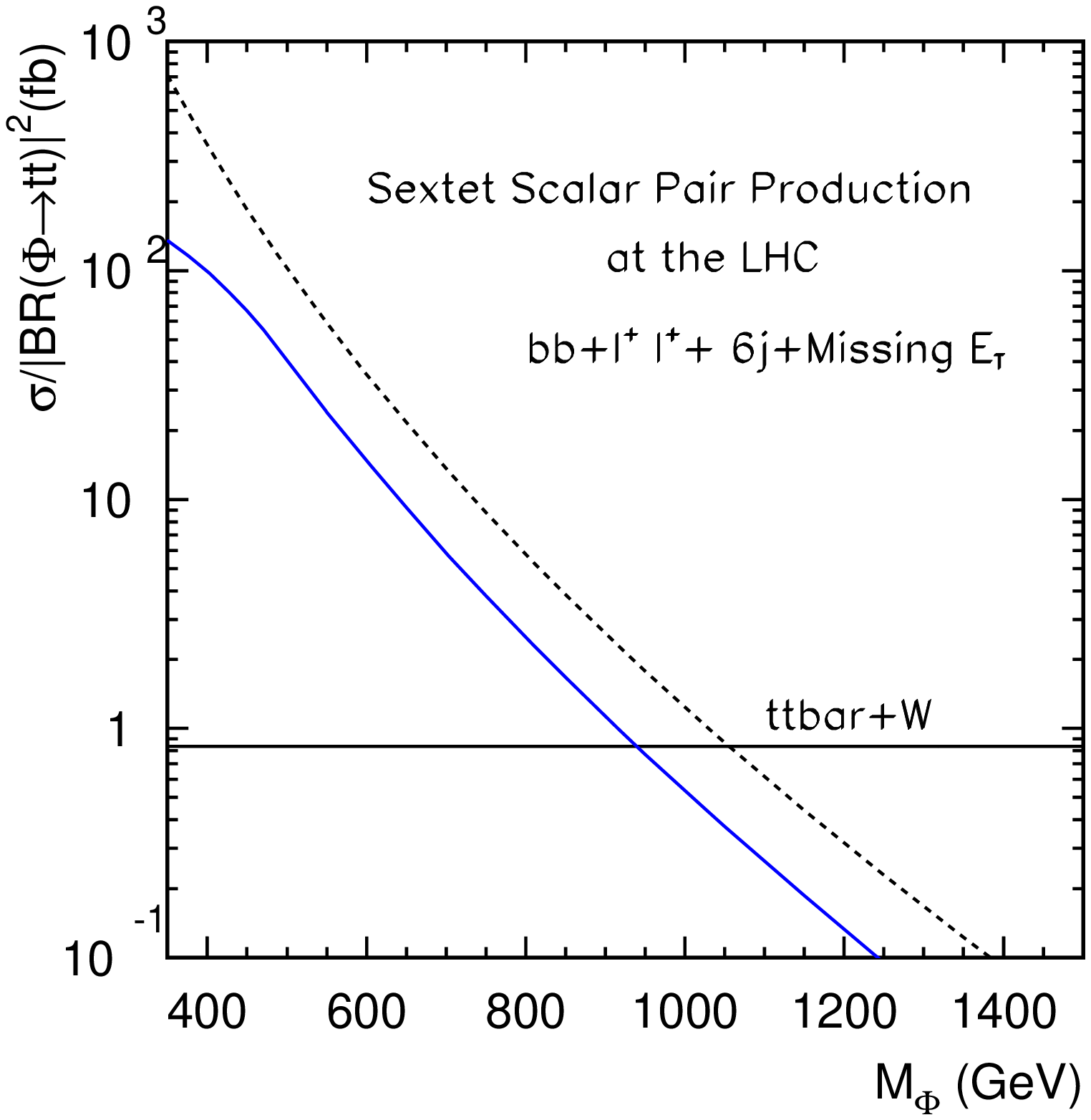}
\includegraphics[scale=1,width=8cm]{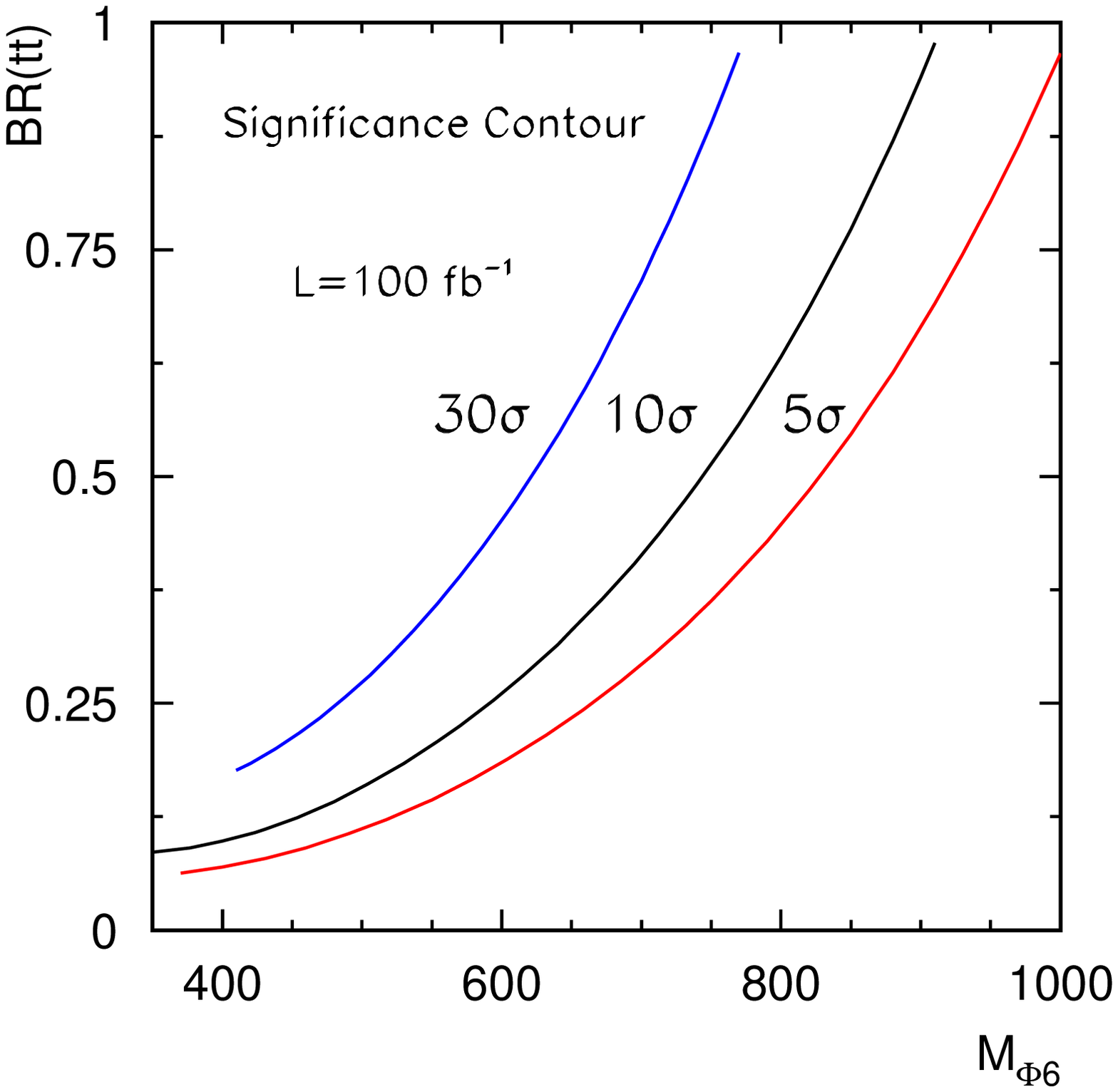}
\caption{Production Rate normalized by BR$(\Phi_6\to tt)^2$ and Significance Contour. Dashed(solid) curves in left plot represent production before(after) selection cuts.}
\label{final}
\end{figure}

Since the production rate of our signal only depends on the mass $M_{\Phi_6}$ and branching ratio of $\Phi_6$ decay to a top quark pair, we scan these two parameters to study the discovery potential. 
We summarize our results in Fig. \ref{final} as the signal production rate for $bb\bar{b}\bar{b}+\ell^\pm\ell^\pm+\cancel{E}_T+4j$ from $\Phi_6 \bar{\Phi}_6$ with
SM $t\bar{t}W^\pm$ background included.
We use a factor of $25\%$ 
in both plots in Fig. \ref{final} for tagging two $b$ jets  with $50\%$ effeciency to tag each $b$-jet. The SM 
background is taken as 1 fb in the significance contour. 
As we can see in the left plot of Fig. \ref{final}, for 100 fb$^{-1}$ luminosity, the statistical significance can 
surpass the $5\sigma$ level for $M_{\Phi_6} \lesssim 800 \rm{GeV}$ if BR($\Phi \to tt$) is about 0.5.
Also note that the mass of the sextet scalar can be determined by reconstructing two hadronically 
decaying top (or anti-top) quarks, and the branching ratio of $\Phi_6\to tt$ can be roughly estimated from the total 
signal event rate if the one can understand the backgorund sufficiently well. 
 No reconstruction selection has been implented since we did not simulate the
events with initial state/final state radiation and the reconstruction efficiency is thus unknown. In principle, we expect that the $S/\sqrt{B}$ can be 
further improved by including reconstruction. 

\section{Conclusion}
In this paper, we discuss the production of a new exotic particle, a color sextet scalar, at the CERN Large Hadron Collider.
Taking a purely phemenological approach, we discuss the discovery of the color sextet scalar through its decay into a top-top quark pair. The
unique feature of same sign dilepton plus multijet makes it easy to identify and reconstruct the color sextet scalar object. Due to its
large QCD production, it is possible to cover the color sextet scalar up to a mass range of 1 TeV for 100 fb$^{-1}$ integrated luminosity.    

In the text, we only consider the case of $M_{\Phi_6} > 2m_t$, where $\Phi_6$ decays into two on-shell top quarks. In the case $2M_{\Phi_6}<m_Z$, there is a possiblity of a $Z$ decaying into a sextet pair, since $\Phi_6$ carries a $U(1)_Y$ charge, which we expect is highly constrained by LEP data. We also expect to find strong constraints from Tevatron data.  For example, for $M_{\Phi_6}$ just above $m_t+m_b$ threshold, the $\bar{\Phi}_6\Phi_6$ signal will directly contribute to $t\bar{t}X$ sample as the offshell top decay products are soft.

\section*{Acknowledgement}
KW would like to thank Rabi Mohapatra for initiating this project, carefully reading the manuscript and providing valuable suggestions.
We would also like to thank Qing-Hong Cao, Kaoru Hagiwara, Tao Han, Hitoshi Murayama, Mihoko Nojiri, Chris Quigg, 
T.T. Yanagida and C.P. Yuan for useful discussion. 
This work is supported by the World Premier International Research Center Initiative (WPI Initiative), MEXT, Japan. 
KW is supported in part by the Project of Knowledge Innovation Program 
(PKIP) of Chinese Academy of Sciences, Grant No. KJCX2.YW.W10 and would 
like to acknowledge the hospitality of Kavli Institute for Theoretical Physics China (KITPC)
while the work was in progress. 

\end{document}